# On the one parameter unit-Lindley distribution and its associated regression model for proportion data

J. Mazucheli[1], A. F. B. Menezes[1] and S. Chakraborty[2]
[1]Department of Statistics, Universidade Estadual de Maringá, DEs, PR, Brazil
[2]Department of Statistics, Dibrugarh University, India

**Abstract**

In this paper considering the transformation $X = \frac{Y}{1+Y}$, where $Y \sim \text{Lindley}(\theta)$, we propose the unit-Lindley distribution and investigate some of its mathematical properties. A important fact associated with this new distribution is that is possible to obtain the analytical expression for bias correction of the maximum likelihood estimator. Moreover, it belongs to the exponential family. This distribution allows us to incorporate covariates directly in the mean and consequently to quantify the influence on the average of the response variable. Finally, a practical application is present and it is shown that our model fits much better than the Beta regression.

# 1 Introduction

In applied statistic, a common issue is to deal with the uncertainty phenomena observed in the bounded in the interval $(0, 1)$. For example, in real life we often encounter measures like proportion or fraction of a certain characteristic, scores of some ability tests, different index, rates, etc., which lie in the interval $(0, 1)$. In such cases continuous distributions with domain $(0, 1)$ are indispensable to probabilistic modeling of the phenomena. The two parameter Beta distribution (or the Pearson type IV distribution) is the most widely used model for such data in practice, mainly because its flexibility (Johnson et al., 1995). Though many distributions have been purposed and studied as alternatives there is still no agreement on preference of a particular model.

In this paper we introduce a single parameter unit-Lindley distribution, derived from a transformation on the Lindley distribution. As far as we know the only other one-parameter distribution in the unit interval is the Topp-Leone distribution (Topp and Leone, 1955). Nevertheless, the Topp-Leone distribution does not posses important properties such as close form expressions for the moments.

The main advantage of the unit-Lindley distribution relies on the fact that practitioners will have a new quite flexible, unimodal one-parameter distribution which posses several crucial properties that other distributions restricted to the interval $(0, 1)$ do not enjoy. For instance, the unit-Lindley distribution has only a single parameter and closed form expressions for cumulative distribution function (c.d.f), quantile function and simple expression for moments unlike the well known Beta distribution (having two parameters, no closed form for c.d.f. and quantile function) and Kumaraswamy distribution (with two parameters, no closed form for moments) distributions. Moreover because of its simple formula for mean the unit-Lindley distribution allows us to directly incorporate the covariates influencing in the mean in order to quantify their average influence on the response variable. This enabled us to present a new bounded regression model as an alternative to the Beta regression model introduced in the statistical literature (Cepeda-Cuervo, 2001; Ferrari and Cribari-Neto, 2004).

We provide a comprehensive account of statistical properties of the proposed distribution along with an application with data from the access of people in households with inadequate water supply and sewage in the cities of Brazil from the Southeast and Northeast regions, to demonstrate that the unit-Lindley regression yields a better fit than the Beta regression model.

The rest of this paper is structured as follows. In Section 2, we start with the model formulation and investigate several features such as moments, incomplete moments, behavior of the cumulative and probability density functions, Lorenz curve and quantile function. Afterwards, parameter estimation by two different methods are discussed in Section 3. A simulation study is conducted in Section 4 to investigate the performance of the proposed estimators. A real life application related to the proportion of people with inadequate water supply and sewage is analyzed in Section 5. We conclude with some discussion and directions for future extensions in Section 6.

## 2 The unit-Lindley distribution

The Lindley distribution was introduced by Lindley (1958) in the context of Bayesian inference. Its probability density function (p.d.f) is specified by:

$$f(y \mid \theta) = \frac{\theta^2}{1+\theta} (1+y) \exp(-\theta y), \qquad y > 0, \theta > 0.$$

The corresponding cumulative distribution function (c.d.f.) is:

$$F(y \mid \theta) = 1 - \left(1 + \frac{\theta y}{1+\theta}\right) \exp(-\theta y). \tag{1}$$

Ghitany et al. (2008) studied the properties of the Lindley distribution and outlined that its mathematical properties are more flexible than those of the exponential distribution.

From (1) using the transformation $X = Y/(1+Y)$ we propose a new distribution with support on the unit-interval. The c.d.f. and the p.d.f. of the resulting distribution are respectively

given by:

$$F(x \mid \theta) = 1 - \left(1 - \frac{\theta x}{(1+\theta)(x-1)}\right) \exp\left(-\frac{\theta x}{1-x}\right), \qquad 0 < x < 1, \theta > 0. \qquad (2)$$

$$f(x \mid \theta) = \frac{\theta^2}{1+\theta}(1-x)^{-3} \exp\left(-\frac{\theta x}{1-x}\right), \qquad 0 < x < 1, \theta > 0. \qquad (3)$$

The first derivative of (3) is:

$$\frac{\mathrm{d}}{\mathrm{d}x} f(x \mid \theta) = \frac{\theta^2 (\theta + 3x - 3)}{(1+\theta)(x-1)^5} \exp\left(-\frac{\theta x}{1-x}\right)$$

which implies that the p.d.f is unimodal with maximum at $X_{\max} = 1 - \frac{\theta}{3}$ for $\theta < 3$ and $X_{\max} = 0$ for $\theta \geqslant 3$. Figure (1) shows the p.d.f. of the unit-Lindley distribution for selected values of $\theta$.

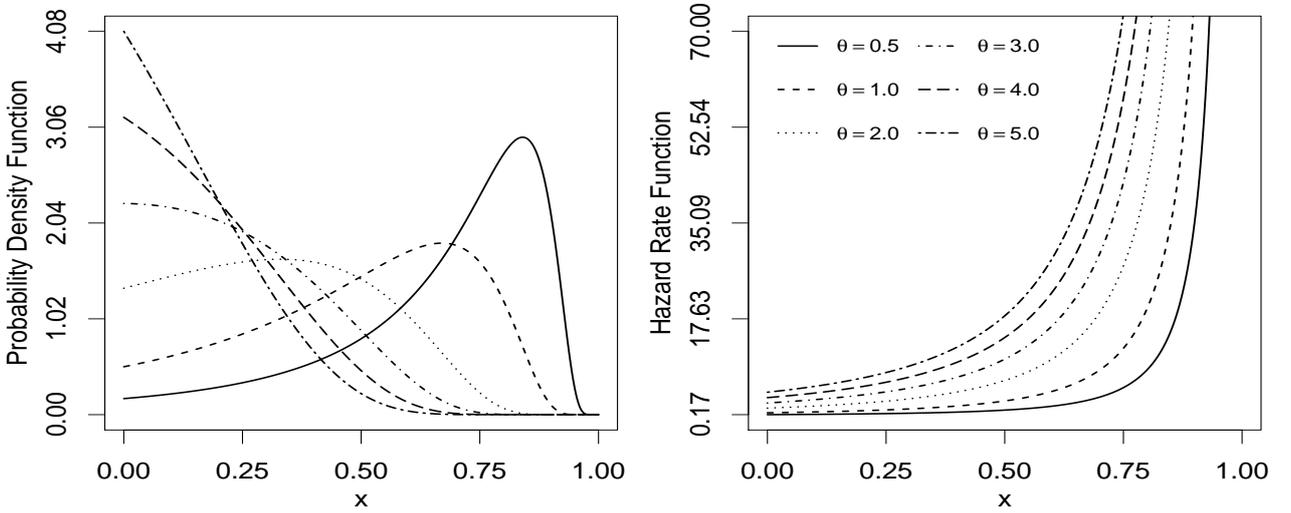

Figure 1: Probability density function and hazard rate function of unit-Lindley distribution for selected values of $\theta$.

In what follows we shall discuss several important statistical properties of the unit-Lindley distribution.

## 2.1 Concavity

**Proposition 1** *The c.d.f. of the unit-Lindley is concave for $\theta > 3$.*

**Proof:** The second derivative of $F(x \mid \theta)$ is

$$F''(x \mid \theta) = \frac{\theta^2 (\theta + 3x - 3)}{(1+\theta)(x-1)^5} \exp\left(-\frac{\theta x}{1-x}\right).$$

This implies for all $x$ in $(0,1)$, $F''(x \mid \theta) > 0$ only if $\theta < 0$ therefore it can never be convex and $F''(x \mid \theta) < 0$ if $\theta > 3$. Hence $F(x)$ is concave function of $x$ for $\theta > 3$. ∎

**Proposition 2** *The c.d.f. of the unit-Lindley is bounded for $\theta > 3/2$ as follows.*

$$x^\theta \leq F(x \mid \theta) \leq 1 - \exp\left(-\frac{\theta x}{1-x}\right).$$

Therefore the c.d.f. can be used to define new premium by distorting survival function (see Wang, 1996).

**Proposition 3** *The p.d.f. of the unit-Lindley is log-concave for all $0 < x < 1$ if $\theta \geq \frac{3}{2}$.*

**Proof:** The second derivative of $F(x \mid \theta)$ is

$$F''(x \mid \theta) = \frac{\theta^2 (\theta + 3x - 3)}{(1+\theta)(x-1)^5} \exp\left(-\frac{\theta x}{1-x}\right).$$

We know that $f(x)$ is log-concave (log-convex) function of $x$ if for all $x$ in $(0,1)$ $\frac{d}{dx} \log f(x \mid \theta)$ is a non-increasing (non-decreasing) function of $x$. Note that

$$\frac{d^2}{dx^2} \log f(x \mid \theta) = \frac{d}{dx} \frac{f'(x)}{f(x)} = \frac{d}{dx} \frac{\theta + 3(x-1)}{(x-1)^2} = \frac{2\theta + 3(x-1)}{(x-1)^3}.$$

This is always $< 0$ for all $x$ in $(0,1)$ when $\theta \geq \frac{3}{2}$. Hence $f(x)$ is log-concave for all $0 < x < 1$, if $\theta \geq \frac{3}{2}$. ∎

As a consequence of this we can state that when $\theta \geq \frac{3}{2}$:

- $f(x)$ is log-concave for all $0 < x < 1$;

- $\int_0^x F(t)\, dt$ is log-concave for all $0 < x < 1$;

- $\bar{F}(x)$ is log-concave for all $0 < x < 1$;

- $\int_x^1 \bar{F}(t)\, dt$ is log-concave for all $0 < x < 1$;

- $\frac{f(x)}{\bar{F}(x)}$ is monotone increasing function in $x$ for all $0 < x < 1$;

- Mean residual life (MRL) is a decreasing function of $x$;

- The distribution is strongly unimodal;

- All moments exist;

- At most has an exponential tail.

## 2.2 Hazard rate function

The hazard rate function of the unit-Lindley distribution is given by:

$$h(x \mid \theta) = \frac{f(x \mid \theta)}{1 - F(x \mid \theta)} = \frac{\theta^2}{(\theta - x + 1)(x - 1)^2}, \qquad 0 < x < 1. \tag{4}$$

Since $\frac{d}{dx} h(x \mid \theta) = \frac{\theta^2}{(x+1)^3(\theta-x+1)^2} [2\theta - 3(x-1)] > 0$ for all $\theta > 0$ the hazard rate function is increasing in $x$. Note that $\lim_{x \to 0} h(x \mid \theta) = \frac{\theta^2}{1+\theta}$ while $\lim_{x \to 1} h(x \mid \theta) = \infty$. The behavior of (4) considering different values of $\theta$ is illustrated on the right side of Figure 1.

## 2.3 Moments

The $k$-th moment about origin of the unit-Lindley distribution is given by:

$$\mu'_k = \mathbb{E}\left(X^k\right) = \frac{k}{(1+\theta)} \int_0^1 \frac{x^{k-1}(1-\theta+x)}{(1-x)} \exp\left(-\frac{\theta x}{1-x}\right) dx, \qquad k = 1, 2, \ldots$$

which can not be solved analytically. In particular, for $k = 1, 2, 3, 4$ we have:

$$\begin{aligned}
\mu'_1 &= \frac{1}{1+\theta}, \\
\mu'_2 &= \frac{1}{1+\theta} \left(\theta^2 e^\theta \mathrm{Ei}(1,\theta) - \theta + 1\right), \\
\mu'_3 &= \frac{1}{1+\theta} \left(e^\theta \mathrm{Ei}(1,\theta) \theta^3 + 3\theta^2 e^\theta \mathrm{Ei}(1,\theta) - \theta^2 - 2\theta + 1\right), \\
\mu'_4 &= \frac{1}{2(1+\theta)} \left(e^\theta \mathrm{Ei}(1,\theta) \theta^4 + 8 e^\theta \mathrm{Ei}(1,\theta) \theta^3 - \theta^3 + 12\theta^2 e^\theta \mathrm{Ei}(1,\theta) - 7\theta^2 - 6\theta + 2\right)
\end{aligned}$$

where $\mathrm{Ei}(a, z) = \int_1^\infty z^{-a} e^{-xz} dx$ is the exponential integral function (Abramowitz and Stegun, 1974).

The higher order moments can be numerically computed using softwares like Mathematica which provide routine for Lambert W function. From the plots of the mean, variance, skewness and kurtosis of the unit-Lindley distribution in Figure 2 it is observed that the mean decreases and skewness increases with the increase in $\theta$ whereas the kurtosis initially decreases then increases with $\theta$.

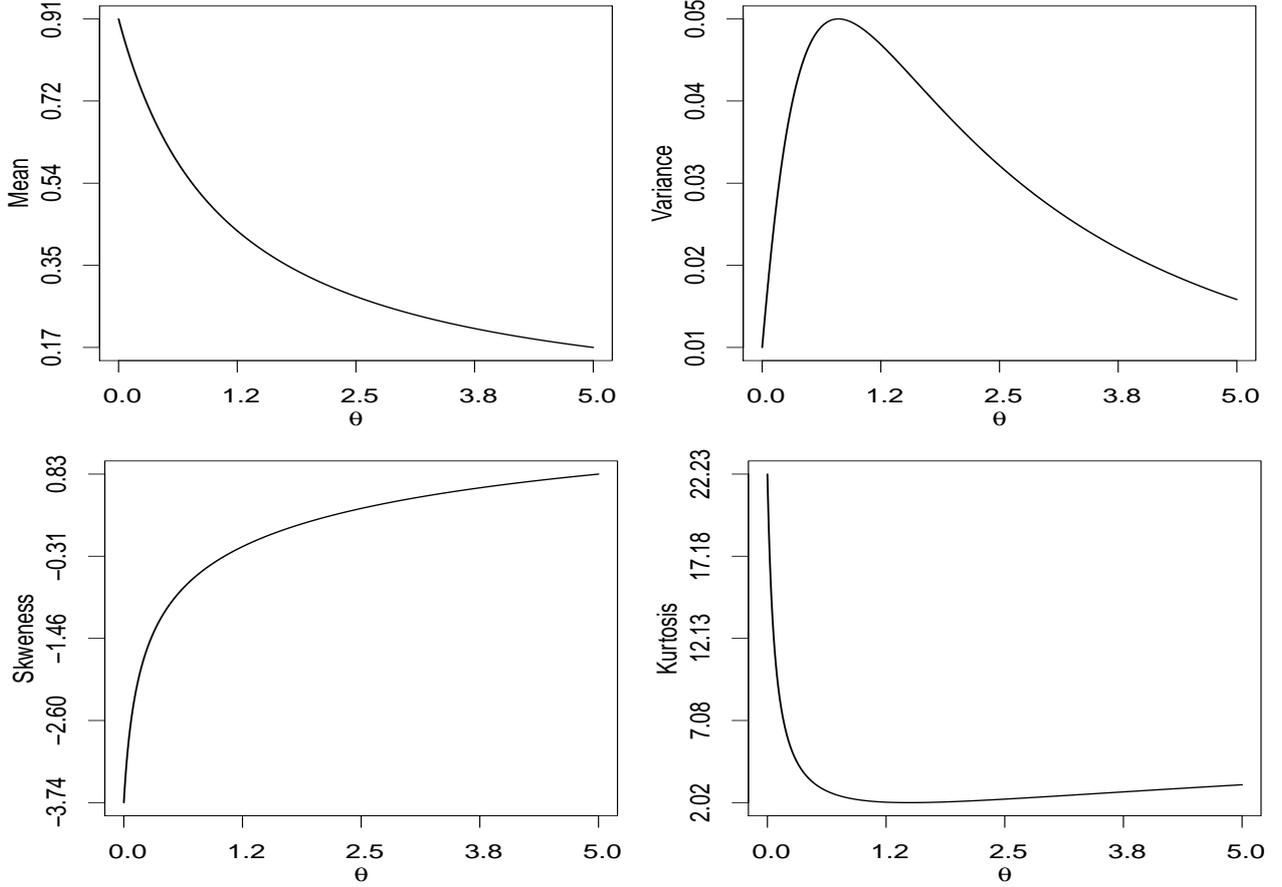

Figure 2: Mean, variance, skewness and kurtosis of unit-Lindley as a function of $\theta$.

## 2.4 Incomplete moments

The $k$-th incomplete moment of the unit-Lindley distribution is given by:

$$T_k(t) = \mathbb{E}\left(X^k \mid x < t\right) = \frac{k}{(1+\theta)} \int_0^t \frac{x^{k-1}(1-\theta+x)}{(1-x)} \exp\left(-\frac{\theta x}{1-x}\right) dx, \qquad k = 1, 2, \ldots$$

which can not be solved analytically. In particular, for $k = 1, 2$ we have:

$$T_1(t) = \frac{[1 + (\theta - 1)t]\, e^{\frac{t-\theta}{t-1}}}{(t-1)(\theta+1)},$$

$$T_2(t) = \frac{e^{\frac{t-\theta}{t-1}} + (t-1)\left[\theta^2 e^\theta (\theta+3)\mathrm{Ei}(1,\theta) - \theta^2 - 2\theta + 1\right] - \theta^2 e^\theta (\theta+3)(t-1)\mathrm{Ei}\left(1, -\theta/(t-1)\right)}{[(2t-1)\theta - t + 1]}$$

## 2.5 Mean residual life function

For a nonnegative continuous random variable $X$ the mean residual life function is defined as $\mu(x) = \mathbb{E}(X - x \mid X > x)$ and is calculated by:

$$\mu(x) = \frac{1}{S(x)} \int_x^\infty S(y) dy.$$

Considering $S(x) = S(x \mid \theta) = \left(1 - \frac{\theta x}{(1+\theta)(x-1)}\right) \exp\left(-\frac{\theta x}{1-x}\right)$, the survival function of the unit-Lindley distribution, we have as result $\mu(x \mid \theta) = \frac{(x-1)^2}{1+\theta+x}$. Note that $\lim_{x \to 0} \mu(x \mid \theta) = \frac{1}{1+\theta}$ while $\lim_{x \to 1} \mu(x \mid \theta) = 0$. The behavior of $\mu(x)$ for different values of $\theta$ is illustrated in Figure 3.

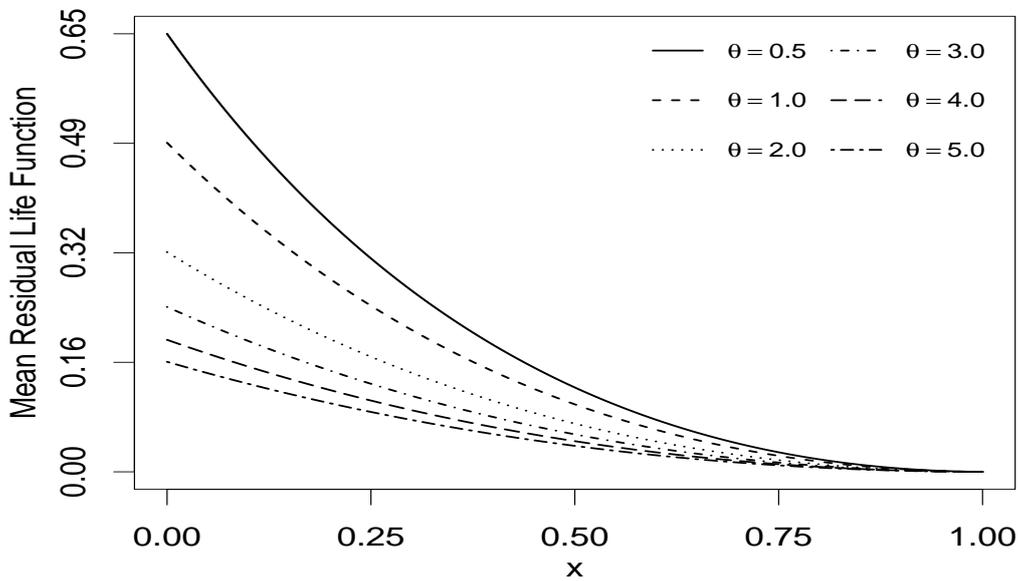

Figure 3: Mean residual life function of the unit-Lindley for selected values of $\theta$.

## 2.6 Mean deviation

As pointed out, for example in Ghitany et al. (2008), the amount of scatter in a population is measured to some extent by the totality of deviations from the mean and the median. These are known as the mean deviation about the mean and the mean deviation about the median, defined as:

$$\delta(X) = \int_x^\infty |X - m| f(x) dx = 2 \left[ mF(m) - \int_0^m x f(x) dx \right], \quad (5)$$

where $m = \mathbb{E}(X)$ or $m = \text{Median}(X)$. Considering (2) and (3) in (5) we have:

$$\delta(X) = \frac{2}{1+\theta} \left( \left[ e^{-\frac{\theta m}{1-m}} (1-m) + m(1+\theta) - 1 \right] \right).$$

For $m = \mathbb{E}(X)$ we get $\delta(X) = \frac{2\theta e^{-1}}{(1+\theta)^2}$. Considering $m = Q(0.5 \mid \theta)$ we have the expression for the mean deviation about the median. The expression for $Q(\cdot \mid \theta)$ is given bellow.

## 2.7 Lorenz curve

The Lorenz curve for a random variable $X$ is defined as:

$$L(F(q)) = \frac{1}{\mathbb{E}(X)} \mathbb{E}(X \mid X \leq q) F(q). \tag{6}$$

For the unit-Lindley distribution we have:

$$\mathbb{E}(X \mid X \leq q) F(q) = \frac{1}{(1+\theta)(q-1)} \left[ e^{-\frac{\theta q}{1-q}} (1 - q + \theta q) + q - 1 \right].$$

From (6) we obtain the Lorenz function for the unit-Lindley distribution written as:

$$L(p) = \frac{1}{(1+\theta)^2 (p-1)} \left[ e^{-\frac{\theta p}{1-p}} (1 - p + \theta p) + p - 1 \right],$$

where $q = F^{-1}(p)$ is given in (7).

## 2.8 Quantile function

Let $X$ be a unit-Lindley random variable with c.d.f (2). The quantile function, $Q(p) = F^{-1}(p)$, can be written as:

$$Q(p \mid \theta) = \frac{1 + \theta + W_{-1}\left((1+\theta)(p-1) e^{-(1+\theta)}\right)}{1 + W_{-1}\left((1+\theta)(p-1) e^{-(1+\theta)}\right)}, \tag{7}$$

such that $0 < p < 1$ and $W_{-1}$ denotes the negative branch of the Lambert $W$ function (Jodra, 2010).

## 2.9 Stress strength reliability

Suppose that $X$ and $Y$ are two independent unit-Lindley random variables with parameters $\theta_1$ and $\theta_2$, respectively, having p.d.f's $f_X(\cdot)$ and $f_Y(\cdot)$. Then the stress-strength reliability measure, (Kotz et al., 2003) is given by:

$$\begin{aligned} R &= P(Y < X) = \int_0^1 f_X(x \mid \theta_1) F_Y(x \mid \theta_2) dx \\ &= \frac{\theta_2^2 \left( \theta_1 \theta_2^2 + 2\theta_1^2 \theta_2 + \theta_1^3 + \theta_2^2 + 4\theta_1 \theta_2 + 3\theta_1^2 + \theta_2 + 3\theta_1 \right)}{(\theta_1 + \theta_2)^3 (1+\theta_2)(1+\theta_1)}. \end{aligned} \tag{8}$$

## 2.10 Exponential Family

A distribution belong to the exponential family if it is of the form (Dobson, 2001)

$$f(x \mid \theta) = \exp\left[Q(\theta) T(x) + D(\theta) + S(x)\right].$$

It can be easily seen that the proposed distribution belongs to the exponential family by rewriting the pdf in equation (3) as

$$f(x \mid \theta) = \exp\left[-\frac{\theta x}{1-x}\right] \exp\left[\log\frac{\theta^2}{1+\theta}\right] \exp\left[\log(1-x)^{-3}\right]$$

where $Q(\theta) = \theta, T(x) = \frac{x}{1-x}, D(\theta) = \log\frac{\theta^2}{1+\theta}, S(x) = \log(1-x)^{-3}$.

Then, $T(\mathbf{x}) = \sum_{i=1}^{n} \frac{x_i}{1-x_i}$ is a complete sufficient estimator for $\theta$ based on a sample of size $n$ from the proposed distribution. Beside that, since the distribution is exponential family a minimum-variance unbiased estimator (MVUE) can be obtained by bias corrected MLE.

# 3 Estimation

In what follows, we shall consider the estimation of parameter $\theta$ of the unit-Lindley distribution by the maximum likelihood methodology and method of moments. For the maximum likelihood estimator of $\theta$ we derive the closed-form expressions for the second order bias-correction.

## 3.1 Maximum likelihood estimator

Let $X_1, \ldots, X_n$ be a random sample from the unit-Lindley distribution with p.d.f. (3). Then, observed $\boldsymbol{x} = (x_1, \ldots, x_n)$, the log-likelihood function of $\theta$ can be written as:

$$\ell(\theta \mid \boldsymbol{x}) \propto 2n \log\theta - n \log(1+\theta) - \theta\, t(\boldsymbol{x}) \tag{9}$$

where $t(\boldsymbol{x}) = \sum_{i=1}^{n} \frac{x_i}{1-x_i}$. The maximum likelihood estimate of $\theta$, $\widehat{\theta}$, is obtained by solving the following linear equation:

$$\frac{d}{d\theta}\ell(\theta \mid \boldsymbol{x}) = \frac{2n}{\theta} - \frac{n}{1+\theta} - t(\boldsymbol{x}) = 0$$

which gives:

$$\widehat{\theta} = \frac{1}{2\, t(\boldsymbol{x})}\left[n - t(\boldsymbol{x}) + \sqrt{t(\boldsymbol{x})^2 + 6 n\, t(\boldsymbol{x}) + n^2}\right]. \tag{10}$$

Next
$$\frac{d^2}{d\theta^2}\ell(\theta \mid \boldsymbol{x}) = \frac{n}{(1+\theta)^2} - \frac{2n}{\theta^2} < 0$$

for all $\theta$, in particular for $\theta = \widehat{\theta}$.

Since $\frac{d^2}{d\theta^2}\ell(\theta \mid \boldsymbol{x})$ is data-independent we have that $n\mathbb{E}\left[\frac{d^2}{d\theta^2}\log f(X \mid \theta)\right] = \frac{d^2}{d\theta^2}\ell(\theta \mid \boldsymbol{x})$. Thus, the expected Fisher information is $\mathbb{I}(\widehat{\theta}) = \frac{2n}{\theta^2} - \frac{n}{(1+\theta)^2}$. The variance of $\widehat{\theta}$ is just the inverse of the expected Fisher information written as: $\mathbb{V}(\widehat{\theta}) = \frac{\theta^2(1+\theta)^2}{n(\theta^2+4\theta+2)}$. It is easy to see that for $\phi = g(\theta) = \mathbb{E}(X)$ we have $\widehat{\phi} = \widehat{\mathbb{E}}(X) = \frac{1}{1+\widehat{\theta}}$ and $\mathbb{V}(\widehat{\phi}) = \frac{\theta^2}{n(\theta^2+4\theta+2)}$.

Cox and Snell (1968) provided a framework for estimating the bias, to $\mathcal{O}(n^{-1})$ for the maximum likelihood estimators of the parameters of regular densities. Hence, subtracting the estimated bias from the original maximum likelihood estimator produces a bias-corrected estimator that is unbiased to $\mathcal{O}(n^{-2})$. Following Cox and Snell (1968) we have the analytical expression for bias-correction of an scalar $\widehat{\theta}$, given by:

$$\mathcal{B}\left(\widehat{\theta}\right) = \left(\kappa^{11}\right)^2 \left[0.5\,\kappa_{111} + \kappa_{11,1}\right] + \mathcal{O}(n^{-2}) \tag{11}$$

where $\kappa^{11} = \mathbb{E}\left[-\frac{d^2}{d\theta^2}\ell(\theta \mid \boldsymbol{x})\right]^{-1} = \frac{\theta^2(1+\theta)^2}{n(\theta^2+4\theta+2)}$, $\kappa_{11,1} = \mathbb{E}\left[-\frac{d^2}{d\theta^2}\ell(\theta \mid \boldsymbol{x}) \times \frac{d}{d\theta}\ell(\theta \mid \boldsymbol{x})\right] = 0$ and $\kappa_{111} = \mathbb{E}\left[-\frac{d^3}{d\theta^3}\ell(\theta \mid \boldsymbol{x})\right] = \frac{2n(\theta^3+6\theta^2+6\theta+2)}{\theta^3(1+\theta)^3}$.

Thus, the bias-corrected maximum likelihood estimator $\widetilde{\theta}$ is:

$$\widetilde{\theta} = \widehat{\theta} - \frac{\widehat{\theta}^5 + 7\widehat{\theta}^4 + 12\widehat{\theta}^3 + 8\widehat{\theta}^2 + 2\widehat{\theta}}{(\widehat{\theta}^2 + 4\widehat{\theta} + 2)^2 n}. \tag{12}$$

where the right hand side is $\widehat{\mathcal{B}}\left(\widehat{\theta}\right)$.

Re-parameterizing (3) in terms of the mean $\mu = \frac{1}{1+\theta}$, such that the maximum likelihood of $\mu$ is

$$\widehat{\mu} = -\frac{1}{2\,t(\boldsymbol{x})}\left[n + t(\boldsymbol{x}) - \sqrt{t(\boldsymbol{x})^2 + 6\,n\,t(\boldsymbol{x}) + n^2}\right],$$

and the bias-corrected maximum likelihood estimator for $\widetilde{\mu}$ is given by

$$\widetilde{\mu} = \widehat{\mu} - \frac{2\,\widehat{\mu}^2\,(2\,\widehat{\mu} - 2)}{n\,(\widehat{\mu}^2 - 2\,\widehat{\mu} - 1)^2}.$$

## 3.2 Method of moment estimator

Let $X_1, \ldots, X_n$ be a random sample from de unit-Lindley distribution with p.d.f (3). The method of moment estimator, $\widehat{\theta}_{MME}$, of $\theta$ is given by:

$$\widehat{\theta}_{MME} = \frac{1 - \overline{X}}{\overline{X}} = \frac{1}{\overline{X}} - 1 \tag{13}$$

which is positively biased, i.e. $E(\widehat{\theta}) - \theta > 0$.

**Proof:** Let $\widehat{\theta}_{MME} = g(\overline{X})$ and $g(t) = \frac{1}{t} - 1$ for $t > 0$. Since $g''(t) = \frac{2}{t^3} > 0$, $g(t)$ is strictly convex. Thus, by Jensen's inequality, we have $E(g(\overline{X})) > g(E(\overline{X}))$. Since $g(E(\overline{X})) = g\left(\frac{1}{1+\theta}\right) = \theta$ we have $E(\widehat{\theta}) > \theta$. ∎

Using the delta method we have that the asymptotic variance of (13) is given by

$$\mathbb{V}(\widehat{\theta}_{MME}) = \frac{1}{\overline{X}^2} \mathbb{V}(\overline{X}), \tag{14}$$

where

$$\mathbb{V}(\overline{X}) = \frac{\theta^2 \, \mathrm{e}^\theta \, \mathrm{Ei}\,(1, \theta) - \theta + 1}{n^2 \, (\theta + 1)}.$$

# 4 Simulation study

In this section, we conduct a Monte Carlo simulation in order to evaluate and compare the finite-sample behavior of the maximum likelihood estimators (MLE), its bias-corrected counterpart obtained by the Cox-Snell methodology (BCE) and the moment estimators (MME) of the parameter $\theta$ that index the unit-Lindley distribution.

We generated samples of size $n = 10, 20, 40, 60$ and $80$ considering $\mathbb{E}(X) = 0.1, 0.2, \ldots, 0.7$, which implies that $\theta = 9.00, 4.00, \ldots, 0.43$. To simulate observations from the unit-Lindley distribution we generated $Y$ from Lindley distribution (see, `rlindley` function in `LindleyR` library) and then use the transformation $X = Y/(1 + Y)$. The simulation experiment was repeated $M = 10.000$. The evaluation was performed based on the estimated bias and root mean-squared error (RMSE).

Table 1 shows that MLE and MME of $\theta$ are positive biased, while the BCE estimator achieve substantial bias reduction, especially for small and moderate sample sizes. It is also observed that the RMSE decreases as $n$ increases, as expected. Additionally, the root mean squared errors of the corrected estimates are smaller than those of the uncorrected estimates.

Table 1: Estimated bias (root mean-squared error) of $\theta$.

| $\theta$ | $n$ | MLE | MME | BCE |
|---|---|---|---|---|
| | 10 | 0.9005 (3.3515) | 0.7898 (3.3314) | 0.0026 (2.9083) |
| | 20 | 0.4240 (2.0844) | 0.3688 (2.0877) | -0.0011 (1.9398) |
| 9.00 | 40 | 0.2077 (1.3864) | 0.1803 (1.3962) | 0.0005 (1.3369) |
| | 60 | 0.1364 (1.1046) | 0.1187 (1.1152) | -0.0005 (1.0781) |
| | 80 | 0.1036 (0.9488) | 0.0904 (0.9588) | 0.0013 (0.9315) |
| | 10 | 0.3634 (1.3720) | 0.2918 (1.3718) | 0.0058 (1.1965) |
| | 20 | 0.1719 (0.8586) | 0.1380 (0.8742) | 0.0026 (0.8013) |
| 4.00 | 40 | 0.0847 (0.5764) | 0.0682 (0.5926) | 0.0022 (0.5566) |
| | 60 | 0.0553 (0.4625) | 0.0438 (0.4770) | 0.0008 (0.4519) |
| | 80 | 0.0400 (0.3953) | 0.0315 (0.4083) | -0.0007 (0.3886) |
| | 10 | 0.1903 (0.7468) | 0.1400 (0.7602) | 0.0027 (0.6571) |
| | 20 | 0.0898 (0.4729) | 0.0658 (0.4931) | 0.0007 (0.4437) |
| 2.33 | 40 | 0.0444 (0.3162) | 0.0328 (0.3347) | 0.0010 (0.3061) |
| | 60 | 0.0291 (0.2533) | 0.0210 (0.2690) | 0.0004 (0.2480) |
| | 80 | 0.0220 (0.2180) | 0.0160 (0.2320) | 0.0006 (0.2145) |
| | 10 | 0.1120 (0.4502) | 0.0770 (0.4722) | 0.0025 (0.3998) |
| | 20 | 0.0536 (0.2877) | 0.0365 (0.3105) | 0.0014 (0.2712) |
| 1.50 | 40 | 0.0263 (0.1943) | 0.0177 (0.2123) | 0.0009 (0.1886) |
| | 60 | 0.0176 (0.1561) | 0.0120 (0.1716) | 0.0008 (0.1530) |
| | 80 | 0.0134 (0.1338) | 0.0092 (0.1477) | 0.0008 (0.1318) |
| | 10 | 0.0648 (0.2847) | 0.0395 (0.3097) | -0.0018 (0.2562) |
| | 20 | 0.0310 (0.1838) | 0.0189 (0.2063) | -0.0008 (0.1744) |
| 1.00 | 40 | 0.0154 (0.1244) | 0.0095 (0.1422) | -0.0002 (0.1212) |
| | 60 | 0.0102 (0.1000) | 0.0063 (0.1152) | -0.0001 (0.0983) |
| | 80 | 0.0075 (0.0861) | 0.0044 (0.0993) | -0.0002 (0.0850) |
| | 10 | 0.0418 (0.1835) | 0.0238 (0.2106) | 0.0005 (0.1665) |
| | 20 | 0.0199 (0.1189) | 0.0115 (0.1412) | 0.0001 (0.1133) |
| 0.67 | 40 | 0.0097 (0.0808) | 0.0056 (0.0975) | 0.0000 (0.0789) |
| | 60 | 0.0063 (0.0650) | 0.0037 (0.0789) | -0.0001 (0.0640) |
| | 80 | 0.0047 (0.0561) | 0.0028 (0.0682) | -0.0001 (0.0554) |
| | 10 | 0.0248 (0.1136) | 0.0122 (0.1388) | 0.0000 (0.1040) |
| | 20 | 0.0116 (0.0743) | 0.0054 (0.0941) | -0.0004 (0.0711) |
| 0.43 | 40 | 0.0057 (0.0507) | 0.0028 (0.0655) | -0.0002 (0.0497) |
| | 60 | 0.0039 (0.0410) | 0.0019 (0.0532) | 0.0000 (0.0404) |
| | 80 | 0.0029 (0.0354) | 0.0014 (0.0459) | 0.0000 (0.0350) |

## 5 Real data analysis

In this section, our interest lies in impose a regression structure for the variable of interest using the unit-Lindley distribution. In regression analysis it is very common to model the mean of the response. Hence, since the unit-Lindley distribution has closed form expression for mean it can be used in this context. It is noteworthy that the p.d.f of unit-Lindley in terms of the mean can be expressed as:

$$f(y \mid \mu) = \frac{(1-\mu)^2}{\mu(1-y)^3} \exp\left(-\frac{y(1-\mu)}{\mu(1-y)}\right) \qquad (15)$$

where $0 < y < 1$ and $0 < \mu < 1$.

Let $Y_1, \ldots, Y_n$ be $n$ independent random variables, where $Y_i \sim \text{UL}(\mu_i)$ for $i = 1, \ldots, n$. The regression model is defined supposing that the mean of $Y_i$ satisfies the following functional relation:

$$g(\mu_i) = \mathbf{x}_i^\top \boldsymbol{\beta} \qquad (16)$$

where $\boldsymbol{\beta} = (\beta_1, \ldots, \beta_p)^\top$ is a $p$-dimensional vector of regression coefficients ($p < n$) and $\mathbf{x}_i^\top = (x_{i1}, \ldots, x_{ip})$ denotes the observations on $p$ known covariates. Note that the variance of $Y_i$ is a function of $\mu_i$ and, as a consequence, of the covariate values, which implies that non-constant response variances are naturally accommodated into the model. We shall assume that the mean link function $g(\cdot)$ is a strictly monotonic and twice differentiable link function that maps $(0, 1)$ into $\mathbb{R}$. Possibilities for this function are the c.d.f.'s of the normal or the logistic distribution, among others (see McCullagh and Nelder, 1989).

Under a classic approach, the unknown parameter vector $\boldsymbol{\beta} = (\beta_1, \ldots, \beta_p)^\top$ are estimated by maximizing the log-likelihood function, which can be expressed as:

$$\ell(\boldsymbol{\beta}) = \sum_{i=1}^{n} \ell_i(\mu_i), \qquad (17)$$

where

$$\ell_i(\mu_i) = 2\log(1-\mu_i) - \log(\mu_i) - 3\log(1-y_i) - \frac{y_i(1-\mu_i)}{\mu_i(1-y_i)}. \qquad (18)$$

The data set used in this section is about the access of people in households with inadequate water supply and sewage in the cities of Brazil from the Southeast and Northeast regions. We are interested in analyzing the association between proportion of households with inadequate water supply and sewage and some sociodemographic variables of these cities. The data are available from `http://atlasbrasil.org.br/2013/`, consist of 3197 cities and all variables were measured during the census in 2010.

Specially, we consider the following covariates: region (REG = 0 for Southeast, REG = 1 for Northeast), life expectancy (LIFE), income per capita (INCPC) and human development index (HDI). We also consider the logit link function, hence the regression structure assume for

$\mu_i$ is given by:

$$\text{logit}(\mu_i) = \beta_0 + \beta_1 \text{HDI}_i + \beta_2 \text{REG}_i + \beta_3 \text{INCPC}_i + \beta_4 \text{LIFE}_i \qquad (19)$$

For sake of comparison we also fit the Beta regression model (Cepeda-Cuervo, 2001; Ferrari and Cribari-Neto, 2004). The procedure NLMIXED (SAS, 2010) was used to perform the required computations.

In Table 2 are presented the estimates, the standard errors and the 95% confidence intervals for the parameters of both models. Although the models under investigation provide the same effect of the covariates under the response variable, it can be seen that the estimates of $\beta_1$ and $\beta_2$ are quite different. Moreover, considering the 95% confidence interval obtain, we can see that all covariates are significant to explain the mean of the response variable. For instance, cities with greater values for HDI tend to have less proportion of households with inadequate water supply and sewage.

Table 2: Summary of the parameters of the fitted models.

| Parameter | Beta | | | unit-Lindley | | |
|---|---|---|---|---|---|---|
| | Estimate | S. E. | 95% C. I. | Estimate | S. E. | 95% C. I. |
| $\beta_0$ | 2.0806 | 0.6230 | (0.8595; 3.3017) | 5.7060 | 0.7831 | (4.1712; 7.2409) |
| $\beta_1$ | -2.8030 | 0.5875 | (-3.9545; -1.6515) | -7.8670 | 0.6239 | (-9.0899; -6.6441) |
| $\beta_2$ | 0.8228 | 0.0475 | (0.7297; 0.9160) | 0.9736 | 0.0510 | (0.8736; 1.0736) |
| $\beta_3$ | -0.0014 | 0.0002 | (-0.0018; -0.0010) | -0.0012 | 0.0001 | (-0.0014; -0.0009) |
| $\beta_4$ | -0.0349 | 0.0098 | (-0.0541; -0.0158) | -0.0471 | 0.0127 | (-0.0719; -0.0223) |
| $\phi$ | 12.7788 | 0.3515 | (12.0898; 13.4678) | — | — | — |

In order to evaluate the fitted models, we calculate the residuals introduced by Cox and Snell (1968). This residuals are defined as:

$$\widehat{e}_i = -\log\left[1 - \widehat{F}(y_i)\right], \qquad i = 1, \ldots, n,$$

where $\widehat{F}(\cdot)$ is an estimated of the cumulative distribution function.

According to Lawless (2003) if the model is appropriate, so the $\widehat{e}_i$ should behave approximately like a sample from the standard exponential distribution. Figure 4 shows the probability-probability (P-P) plots, where the empirical probabilities of $\widehat{e}_i$ are compared with the standard exponential distribution. It is observed that the plotted points for the unit-Lindley regression are most closest to the diagonal line.

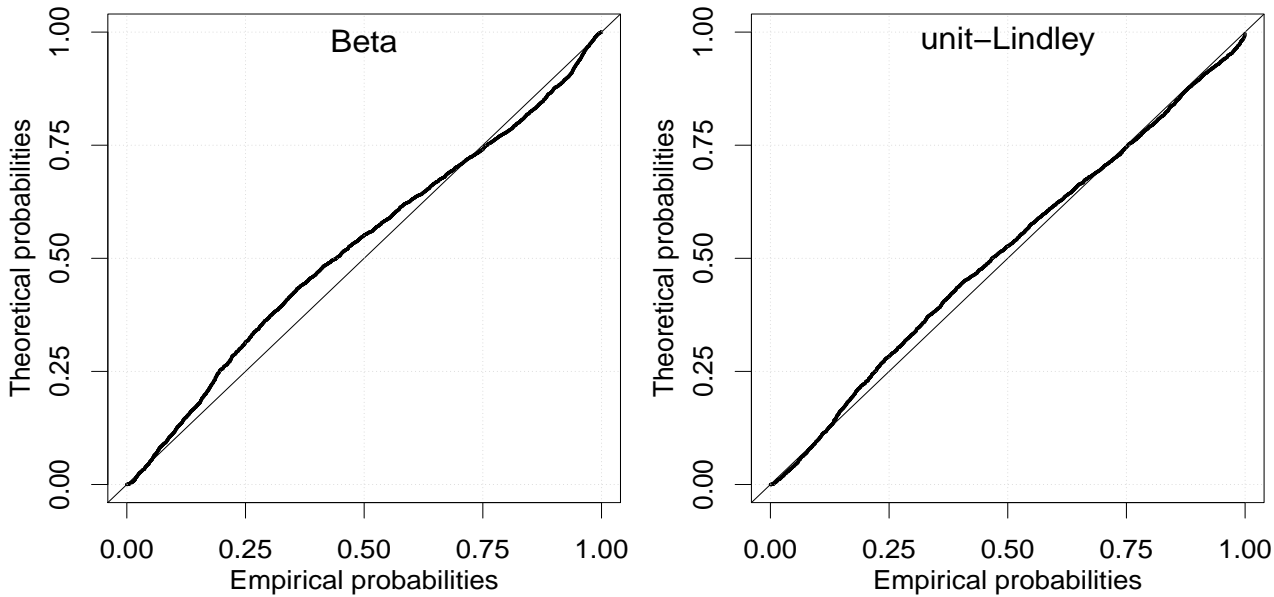

Figure 4: Theoretical and empirical probabilities of the Cox-Snell residuals.

Finally, to discriminate between the unit-Lindley and Beta regression we compute the likelihood-based statistics (Akaike's Information Criterion (AIC), Bayesian Information criterion (BIC) and Hannan-Quinn Information Criterion (HQIC)). Furthermore, we consider the generalized likelihood statistic introduced by Vuong (1989) for comparison of non-nested models, in order to try to choose the best regression model. Based on the results from Table 3 we can conclude that the unit-Lindley regression provide the better fit.

Table 3: Likelihood-based statistics.

| Model | AIC | BIC | HQIC | Voung ($p$-value) |
|---|---|---|---|---|
| unit-Lindley | -11470.6765 | -11440.3266 | -11459.7950 | 5.5654 ($< 0.0000$) |
| Beta | -11038.7573 | -11002.3375 | -11025.6996 | |

# 6  Concluding remarks

In many fields of applied science certain indicators, percentages, proportions, ratios and rates measured in $(0,1)$ scale have been treated as study variables for characterization of distinct phenomena. The current statistical literature provide very few choices of models to deal with such variables. Two main such distributions are the Beta and Kumaraswamy distributions. The present paper has contributed a new one parameter probability distribution with bounded domain constructed by an simple intuitive variable transformation in the Lindley distribution. Random sample for the distribution can be easily simulated by simple transformation of sample generated from Lindley distribution. Several statistical properties of the unit-Lindley have been studied. Method of moments and maximum likelihood estimation are discussed and analytical expression for the bias correction of the maximum likelihood estimator are derived. The fact that the unit-Lindley distribution allows us to incorporate a regression structure in the mean of the response variables, allowed it to seen as an alternative which is more parsimonious compared to the Beta regression model. Application of the proposed model to a real data set yields a better fit than the Beta regression model. As such we envisage that our new distribution will be highly utilized across all relevant fields of science.